\documentclass[traditabstract]{aa}
\usepackage{txfonts}
\usepackage{graphicx}
\usepackage{natbib}
\bibpunct{(}{)} {;}{a}{ }{,}
\begin{document}

\title{ALMA reveals a chemically evolved submillimeter galaxy at z=4.76}

\authorrunning{Nagao et al.}
\titlerunning{ALMA reveals a chemically evolved SMG at z=4.76}

\author{
          T. Nagao\inst{1,2}
          \and
          R. Maiolino\inst{3}
          \and
          C. De Breuck\inst{4}
          \and
          P. Caselli\inst{5}
          \and
          B. Hatsukade\inst{2} 
          \and
          K. Saigo\inst{6}
          }
\institute{
   The Hakubi Center for Advanced Research, Kyoto University, 
   Kyoto 606-8302, Japan;
   tohru@kusastro.kyoto-u.ac.jp
   \and
   Department of Astronomy, Kyoto University, 
   Kitashirakawa-Oiwake-cho, Sakyo-ku, Kyoto 606-8502, Japan
   \and
   Cavendish Laboratory, Univerisy of Cambridge, 19 J. J. Thomson Ave., 
   Cambridge CB3 0HE, United Kingdom
   \and
   European Southern Observatory,
   Karl Schwarzschild Strasse 2, 85748 Garching, Germany
   \and
   School of Physics and Astronomy, University of Leeds,
   Leeds LS2 9JT, United Kingdom
   \and
   East Asian ALMA Regional Center,
   National Astronomical Observatory of Japan,
   Osawa 2-21-1, Mitaka, Tokyo 181-8588, Japan
   }
\date{Received xxx; accepted xxx}

\abstract{
   The chemical properties of high-$z$ galaxies provide important information
   for constraining galaxy evolutionary scenarios. However, widely used metallicity 
   diagnostics based on rest-frame optical emission lines are unusable for 
   heavily dust-enshrouded galaxies (such as submillimeter galaxies; SMGs), 
   especially at $z>3$. Here we focus on the flux ratio of the far-infrared 
   fine-structure emission lines [N~{\sc ii}]\,205\,$\mu$m and 
   [C~{\sc ii}]\,158\,$\mu$m to assess the metallicity of high-$z$ SMGs. Through 
   ALMA cycle 0 observations, we have detected the [N~{\sc ii}]\,205\,$\mu$m 
   emission in a strongly [C~{\sc ii}]-emitting SMG, LESS J033229.4--275619 at 
   $z=4.76$. The velocity-integrated [N~{\sc ii}]/[C~{\sc ii}] flux ratio is 
   $0.043 \pm 0.008$. This is the first measurement of the [N~{\sc ii}]/[C~{\sc ii}] 
   flux ratio in high-$z$ galaxies, and the inferred flux ratio is similar to the ratio 
   observed in the nearby universe ($\sim 0.02-0.07$). The velocity-integrated 
   flux ratio and photoionization models suggest that the metallicity in this SMG 
   is consistent with solar, implying that the chemical evolution has progressed 
   very rapidly in this system at $z=4.76$. We also obtain a tight upper limit on the 
   CO(12-11) transition, which translates into CO(12-11)/CO(2-1) $<3.8$ 
   (3$\sigma$). This suggests that the molecular gas clouds in 
   LESS J033229.4--275619 are not significantly affected by the radiation field 
   emitted by the AGN in this system.
}
\maketitle

\section{Introduction}

Investigating the metal content in galaxies is a powerful diagnostic for testing 
galaxy evolutionary models, because the metallicity is determined by the past 
star-formation history, gas inflow, and outflow phenomena. The gas 
metallicity in galaxies has been investigated in galaxies up to z$\sim$3
\citep{2008A&A...488..463M,2010MNRAS.408.2115M}. However, 
rest-frame optical metallicity diagnostics \citep[e.g.,][]{2006A&A...459...85N} 
are not applicable for galaxies at $z>4$, where the optical emission lines 
required to measure the metallicity are redshifted out of the near-infrared 
atmospheric window \citep[see][]{2008A&A...488..463M}. Moreover, 
high-$z$ young galaxies with vigorous star formation are often obscured 
by dust and recognized as, e.g., ultraluminous infrared galaxies (ULIRGs) 
and submillimeter galaxies (SMGs). In these galaxies, the rest-frame optical 
lines are significantly affected by the dust extinction and accordingly the 
optical metallicity diagnostics may lead to large systematic errors. 
For instance, it has been claimed that ULIRGs show systematically lower 
metallicities than expected from the mass-metallicity relation 
\citep{2008ApJ...674..172R,2008ApJ...680..939C}; but this may 
simply be a consequence of the fact that in these heavily obscured 
systems the optical metallicity tracers only probe the outer, less enriched 
regions. \citet{2010A&A...518L.154S} reported that the high dust mass
measured in these galaxies disagrees with the metallicity inferred 
from the optical metallicity diagnostics, suggesting that the optical spectral 
indices are likely unreliable in these dusty systems.

Metallicity diagnostics exploiting far-infrared fine-structure emission lines are
powerful alternatives \citep{2011A&A...526A.149N}. Here we focus on the 
two strongest fine-structure lines at $\lambda_{\rm rest} > 150~\mu$m, i.e.,
[C~{\sc ii}]\,158\,$\mu$m and [N~{\sc ii}]\,205\,$\mu$m. The 
[C~{\sc ii}]\,158\,$\mu$m emission ($\nu_{\rm rest} = 1900.539$ GHz) is one 
of the strongest lines among the whole electromagnetic spectrum and is
consequently observed for some high-$z$ objects even up to $z \sim 6-7$ 
\citep{2005A&A...440L..51M,2009Natur.457..699W,2012arXiv1203.5844V}.
This emission arises mainly in H~{\sc ii} regions and photodissociation 
regions (PDRs), and the relative contribution from those two regions 
depends on the physical conditions of gas clouds \citep[e.g.,][]
{2005ApJS..161...65A}. The [N~{\sc ii}]\,205\,$\mu$m emission 
($\nu_{\rm rest} = 1461.132$ GHz; \citealt{1994ApJ...428L..37B}) arises in 
H~{\sc ii} regions \citep[see][]{2011A&A...526A.149N}. In H~{\sc ii} regions, 
the [N~{\sc ii}]/[C~{\sc ii}] flux ratio depends mostly on the N/C elemental 
abundance ratio, while it is relatively insensitive to other gas physical 
conditions such as the gas density, since their critical densities are similar 
($n_{\rm cr} = 44$ cm$^{-3}$ and 46 cm$^{-3}$ for [N~{\sc ii}]\,205\,$\mu$m 
and [C~{\sc ii}]\,158\,$\mu$m at 8000 K, respectively). Note that the N/C 
ratio is proportional to $Z_{\rm gas}$ at least for 
$\log Z_{\rm gas}/Z_\odot > -0.5$ as observed in Galactic H~{\sc ii} regions 
\citep[e.g.,][]{1998ApJ...497L...1V} because nitrogen is a secondary element 
\citep[see also][]{1999ARA&A..37..487H}. Since the volume ratio of 
H~{\sc ii} regions and PDRs also depends on cloud properties such as 
the gas density and ionization structure (hence the ionization parameter),
the observed [N~{\sc ii}]/[C~{\sc ii}] flux ratio accordingly depends on those 
parameters. Recently, [N~{\sc ii}]\,205\,$\mu$m detections have been 
obtained for three strongly lensed galaxies at z$\sim$4--5
\citep{2012arXiv1203.6852D,2012A&A...538L...4C}; however a
[C~{\sc ii}]\,158\,$\mu$m detection has not been reported in any of these 
galaxies.

Here we show that the [N~{\sc ii}]/[C~{\sc ii}] flux ratio is a good metallicity 
indicator based on photoionization models, and apply this new method to a 
high-$z$ luminous SMG, LESS J033229.4--275619 (hereafter LESS J0332). 
We selected this SMG because this object is starburst-dominated and shows 
intense [C~{\sc ii}]\,158\,$\mu$m emission \citep{2011A&A...530L...8D}, 
making this SMG a good target for the [N~{\sc ii}]\,205\,$\mu$m observation. 
In this $Letter$ we report a clear [N~{\sc ii}]\,205\,$\mu$m detection based on 
our ALMA cycle 0 observation, and discuss the chemical property of a 
high-$z$ SMG. Throughout this $Letter$, we adopt a cosmology with 
$H_0 = 70$ km s$^{-1}$ Mpc$^{-1}$, $\Omega_{\rm m} = 0.27$, and 
$\Omega_\Lambda = 0.73$.

\section{Observations and results}

\begin{figure}
\includegraphics[width=7.4cm]{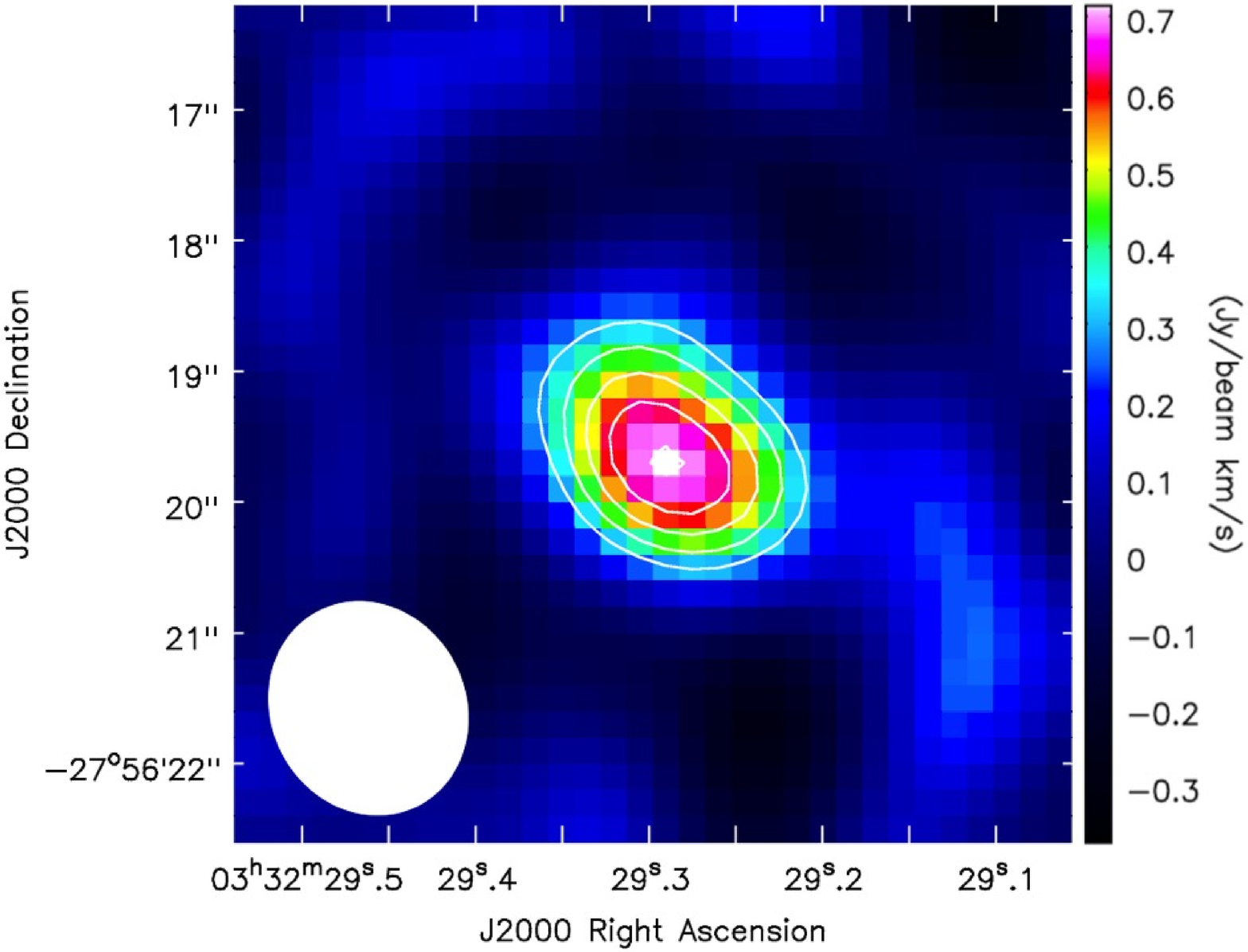}
\caption{
  Velocity-integrated [N~{\sc ii}] map of LESS J0332 after the continuum 
  subtraction, with the spatial sampling of 0.2 arcsec/pixel. The velocity range 
  from --558 km s$^{-1}$ to --154 km s$^{-1}$ (with respect to the Ly$\alpha$ 
  emission-line peak) is integrated. Contours at 3$\sigma$, 4$\sigma$, 
  5$\sigma$, 6$\sigma$, and 7$\sigma$ levels are also given in the map. The
  shape of the synthesized beam is shown at the lower left-hand corner. 
}
\end{figure}

We observed LESS J0332 at $z$ = 4.76, with the band 6 
receiver on the Atacama Large Millimeter/submillimeter Array (ALMA) in the 
dual-polarization setup, as a cycle 0 program. The observation was 
carried out in three separated runs; on 2011 October 1, 2012 January 12, 
and 2012 January 13. Each observing run consisted of 1.2 hours and the 
total observing time is 3.6 hours, including overheads. The receiver was tuned 
to 253.96001 GHz to cover the redshifted [N~{\sc ii}]\,205\,$\mu$m emission 
with the upper sideband, whose redshift is based on our previous 
[C~{\sc ii}]\,158\,$\mu$m detection \citep{2011A&A...530L...8D}. The lower 
sideband was used to cover the redshifted CO(12-11) emission 
($\nu_{\rm rest} = 1381.995$ GHz). The correlator was used in the frequency 
domain mode with a bandwidth of 1875 MHz (488.28 kHz $\times$ 3840 
channels). The observation was performed with 18 antennas in the compact 
configuration, but the data obtained with one antenna was flagged out in the 
last run due to its high system temperature. Callisto was also observed as a 
flux calibrator. The bandpass and phase were calibrated with J0522--364 
and J0403--360, respectively. The atmospheric condition was PWV = 3.0--5.0 
mm in the first run and PWV = 1.0--2.0 mm for the last two runs. 

The data were processed with Common Astronomy Software Applications
(CASA; \citealt{2007ASPC..376..127M,2012arXiv1201.3454P}) in a standard 
manner. A 70-channel (or equivalently, 40.54 km s$^{-1}$) binning was 
applied to the data cube, and then the clean process was applied with the 
natural weighting, which gives a final synthesized beam size of 
$1.67^{\prime\prime} \times 1.48^{\prime\prime}$ (position angle = 26.8 
degree). The [N~{\sc ii}]\,205\,$\mu$m emission of LESS J0332 is clearly 
detected in the continuum-subtracted binned channel map, as shown in 
Fig.~1. 

\begin{figure}
\includegraphics[width=7.5cm]{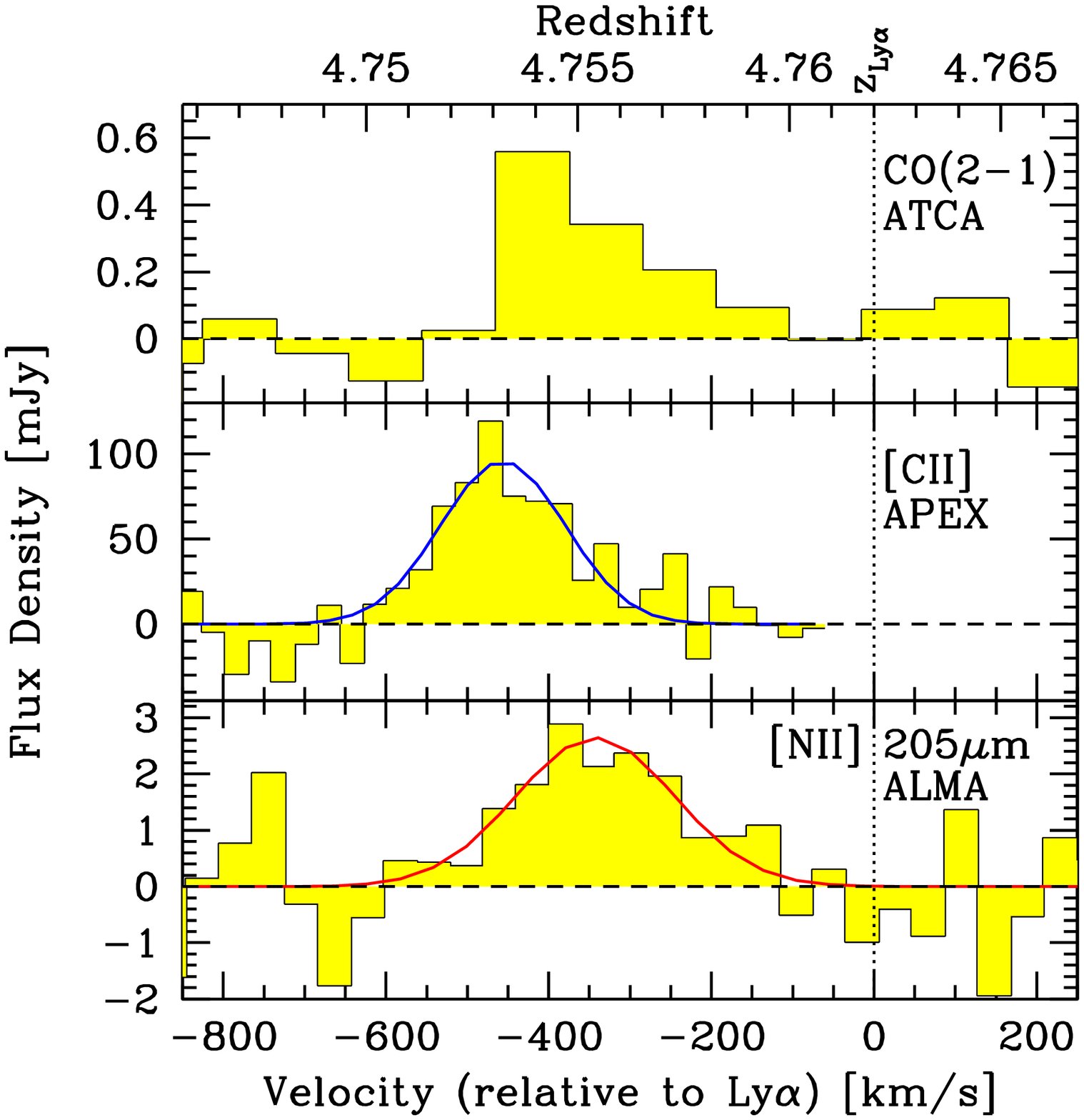}
\caption{
  {\it Top panel:} ATCA CO(2-1) spectrum of LESS J0332 adopting a
  90 km s$^{-1}$ binning \citep{2010MNRAS.407L.103C}.
  {\it Middle panel:} APEX [C~{\sc ii}]\,158\,$\mu$m spectrum with a 28 
  km s$^{-1}$ binning \citep{2011A&A...530L...8D}.
  {\it Bottom panel:} ALMA [N~{\sc ii}]\,205\,$\mu$m spectrum with a 41
  km s$^{-1}$ binning.
  All the spectrum is continuum-subtracted.
  The best-fit Gaussian profile is overlaid on the [C~{\sc ii}] and [N~{\sc ii}]
  spectra. The velocity is relative to the Ly$\alpha$ redshift, and
  the zero velocity is indicated by the dotted line.
}
\end{figure}

Fig.~2 shows the [N~{\sc ii}]\,205\,$\mu$m spectrum 
of LESS J0332 adopting an aperture size of 4.8 arcsec$^2$, with the 
previously reported spectra of CO(2-1) \citep{2010MNRAS.407L.103C} 
and [C~{\sc ii}]\,158\,$\mu$m \citep{2011A&A...530L...8D}. The 
[N~{\sc ii}]\,205\,$\mu$m detection significance is $\sim$8$\sigma$. The 
emission line is fitted with a single Gaussian profile. The best-fit result is 
parameterized by the peak frequency of 253.868$\pm$0.007 GHz, peak 
intensity of 2.58 $\pm$ 0.21 mJy, FWHM of 230 $\pm$ 22 km s$^{-1}$, and 
the velocity-integrated flux of 630 $\pm$ 78 mJy km s$^{-1}$. The redshift 
inferred from the observed [N~{\sc ii}]\,205\,$\mu$m frequency is 
$z_{\rm [NII]} = 4.7555 \pm 0.0002$, which is significantly blueshifted from 
the Ly$\alpha$ redshift. This blueshift is probably caused by the Ly$\alpha$ 
absorption by the intergalactic neutral hydrogen. 
The redshift is consistent with the CO(2-1) redshift, though there is a velocity 
offset, significant at $\sim$$2\sigma$, between the [N~{\sc ii}]\,205\,$\mu$m and 
[C~{\sc ii}]\,158\,$\mu$m lines. This velocity offset may be a consequence of
the modest signal-to-noise ratio in the two spectra, especially in the [C~{\sc ii}] spectrum.
Forthcoming [C~{\sc ii}] ALMA observations of the same objects will either 
confirm or reconcile the velocity discrepancy. However, if this velocity offset 
is confirmed, this results in different [N~{\sc ii}]/[C~{\sc ii}] flux ratios in the 
blue and red components (see \S 3.2).

The redshifted CO(12-11) emission is not detected in our ALMA data. The 
inferred 3$\sigma$ upper limit on the velocity-integrated flux is 344 mJy km 
s$^{-1}$ (adopting the same aperture size as adopted for the 
[N~{\sc ii}]\,205\,$\mu$m line and assuming the same velocity profile).

The continuum emission at 
$\lambda_{\rm obs} \sim 1.2$ mm (i.e., $\lambda_{\rm rest} \sim  210~\mu$m)
is clearly detected with high accuracy. Its flux is $3.5 \pm 0.1$ mJy with the 
same aperture as adopted for the [N~{\sc ii}]\,205\,$\mu$m measurement. 
These measurements were made by combining the upper and lower sideband 
data and excluding the channels affected by the [N~{\sc ii}]\,205\,$\mu$m 
emission and by a relatively high noise level at the edges of the sidebands.

Both the [N~{\sc ii}]\,205\,$\mu$m emission and the continuum emission
are unresolved at our angular resolution. 
The two-dimensional Gaussian fit on the velocity-integrated images results 
in the major and minor axis FWHMs of $1.86 \pm 0.44$ arcsec and 
$1.31 \pm 0.47$ arcsec for the [N~{\sc ii}]\,205\,$\mu$m emission, and 
$1.72 \pm 0.19$ arcsec and $1.55 \pm 0.21$ arcsec for the continuum 
emission. Both results are consistent with the synthesized-beam image shape.

\begin{table}
\caption{Observed properties of LESS J033229.4--275619}
\centering
\begin{tabular}{lll} 
  \hline\hline
  Parameter & Value & Reference 
  \\  \hline  
  $z_{\rm Ly\alpha}$		& 4.762$\pm$0.002			& \citet{2009MNRAS.395.1905C}\\
  $z_{\rm CO(2-1)}$		& 4.755$\pm$0.001			& \citet{2010MNRAS.407L.103C}\\
  $z_{\rm [CII]}$			& 4.7534$\pm$0.0009		& De Breuck et al. (2011)\\
  $z_{\rm [NII]}$			& 4.7555$\pm$0.0002		& this work\\
  $I_{\rm CO(2-1)}$		& 0.09$\pm$0.02 Jy km s$^{-1}$&\citet{2010MNRAS.407L.103C}\\
  $I_{\rm [CII]}$			& 14.7$\pm$2.2 Jy km s$^{-1}$	& De Breuck et al. (2011)\\
  $I_{\rm [NII]}$			& 0.630$\pm$0.078 Jy km s$^{-1}$& this work\\
  $I_{\rm CO(12-11)}$ 	& $<$ 0.344 Jy km s$^{-1}$ (3$\sigma$)& this paper\\
  $\Delta V_{\rm CO(2-1)}$&  160$\pm$65 km s$^{-1}$	& \citet{2010MNRAS.407L.103C}\\
  $\Delta V_{\rm [CII]}$	&  161$\pm$45 km s$^{-1}$	& De Breuck et al. (2011)\\
  $\Delta V_{\rm [NII]}$	&  230$\pm$22 km s$^{-1}$	& this work\\
  \hline
\end{tabular}
\end{table}
 
\section{Discussion}

\subsection{Possible AGN contribution}

\citet{2011ApJ...730L..28G} reported the presence of a Compton-thick 
active galactic nucleus (AGN) in LESS J0332. If the AGN contributes
significantly to the observed far-IR emission lines, its 
interpretation would become accordingly more complex. Within this 
context the CO spectral line energy distribution can help, since it is 
sensitive to the heating energy source. CO lines at high excitation levels 
are significantly stronger when the molecular gas clouds are  affected 
by the X-ray emission from AGNs than in cases without AGNs 
\citep[see, e.g.,][]{2008ApJ...678L...5S}. The nearby ULIRG-Quasar 
Mrk 231 shows strong high-$J$ CO lines up to $J=13-12$, which are 
properly accounted for by introducing X-ray dominated regions (XDRs) 
into models \citep{2010A&A...518L..42V}. At high-$z$ the 
gravitationally magnified quasar at $z=3.91$, APM 08279+5255 also 
shows strong high-$J$ CO lines \citep{2007A&A...467..955W,
2011ApJ...741L..37B}, which are also well described by XDR models. 
The CO spectral line energy distribution of other star-formation-dominated 
high-$z$ galaxies, such as SMM J16359+6612 at $z=2.5$ 
\citep{2005A&A...440L..45W} and IRAS F10214+4724 at $z=2.3$ 
\citep{2008A&A...491..747A}, is completely different from that of APM 
08279+5255, showing weaker high-$J$ CO lines 
\citep[see, e.g., Fig.~14 in][]{2007A&A...467..955W}. 

By combining our upper limit on the CO(12-11) flux and the previous
measurement of the CO(2-1) flux \citep{2010MNRAS.407L.103C}, we 
obtain a 3$\sigma$ upper limit on the flux ratio of CO(12-11)/CO(2-1) 
of 3.8. This upper limit is inconsistent with the CO spectral line energy 
distribution of the quasar APM 08279+5255, but is fully consistent with 
other star-formation-dominated high-$z$ objects \citep[see Fig.~14 in]
[]{2007A&A...467..955W}. This suggests that the molecular 
clouds in LESS J0332 are not described by XDR models, i.e., the AGN 
contribution to the heating and excitation of the ISM in LESS J0332 is 
not significant. This is consistent with our earlier study on LESS J0332 
\citep{2011A&A...530L...8D}, where we estimated that the XDR 
contribution to the [C~{\sc ii}]\,158\,$\mu$m is $\sim$1.3\%, based on the 
absorption-corrected X-ray luminosity of $L_{\rm 2-10 keV} = 2.5 \times 
10^{44}$ erg s$^{-1}$ \citep{2011ApJ...730L..28G} and  a scaling 
relation of $L_{\rm [CII], AGN} = 2 \times 10^{-3} L_{\rm 2-10 keV}$
\citep{2010ApJ...724..957S}.

\subsection{Gas metallicity}

Based on our [N~{\sc ii}]\,205\,$\mu$m detection and our previous 
[C~{\sc ii}]\,158\,$\mu$m detection \citep{2011A&A...530L...8D} in LESS 
J0332, the velocity-integrated flux ratio of [N~{\sc ii}]/[C~{\sc ii}] is inferred 
to be $0.043 \pm 0.008$. Unfortunately, there are only few previous 
measurements on the [N~{\sc ii}]\,205\,$\mu$m line in galaxies (mostly
because this line was not covered by the ISO/LWS wavelength 
range). In the nearby universe, the flux ratio of [N~{\sc ii}]/[C~{\sc ii}] is 
reported only for M82 ($\sim$0.050; \citealt{1994ApJ...427L..17P}), 
Mrk 231 ($\sim$0.067; \citealt{2010A&A...518L..41F}), NGC 1097 
($\sim$0.017; \citealt{2010A&A...518L..60B}), and Arp 220 ($\sim$0.059;
\citealt{2011ApJ...743...94R}). Therefore the [N~{\sc ii}]/[C~{\sc ii}] flux 
ratio of LESS J0332 is similar to the observed ratios reported for nearby 
galaxies, suggesting similar $Z_{\rm gas}$.

At high-$z$ [N~{\sc ii}]\,205\,$\mu$m has been detected in HLS 
J091828.6+514223 at $z=5.24$ \citep{2012A&A...538L...4C}, APM 
08279+5255 at $z=3.91$, and MM 18423+5938 at $z=3.93$ 
\citep{2012arXiv1203.6852D}. However, their [C~{\sc ii}]\,158\,$\mu$m 
line has not been observed and consequently their [N~{\sc ii}]/[C~{\sc ii}] ratio is 
unknown (see \citealt{2009ApJ...691L...1W} and references therein).
Therefore our [N~{\sc ii}]\,205\,$\mu$m detection allows us to infer the 
first measurement of the diagnostic [N~{\sc ii}]/[C~{\sc ii}] flux ratio at 
high-$z$. Note that there are many [N~{\sc ii}]\,122\,$\mu$m detections in
nearby galaxies \citep[e.g.,][]{2011ApJ...728L...7G} and 
also in a few high-$z$ galaxies 
\citep{2011ApJ...740L..29F}. Although there are attempts to infer the 
[N~{\sc ii}]\,205\,$\mu$m flux from the [N~{\sc ii}]\,122\,$\mu$m emission
\citep[e.g.,][]{2012arXiv1203.6852D}, this method may introduce a large 
systematic error because the flux ratio of [N~{\sc ii}]\,122\,$\mu$m and 
[N~{\sc ii}]\,205\,$\mu$m varies by a factor of $\sim$10, i.e., it is strongly 
dependent on the gas density \citep[see][]{2006ApJ...652L.125O}.

\begin{figure}
\includegraphics[width=7.9cm]{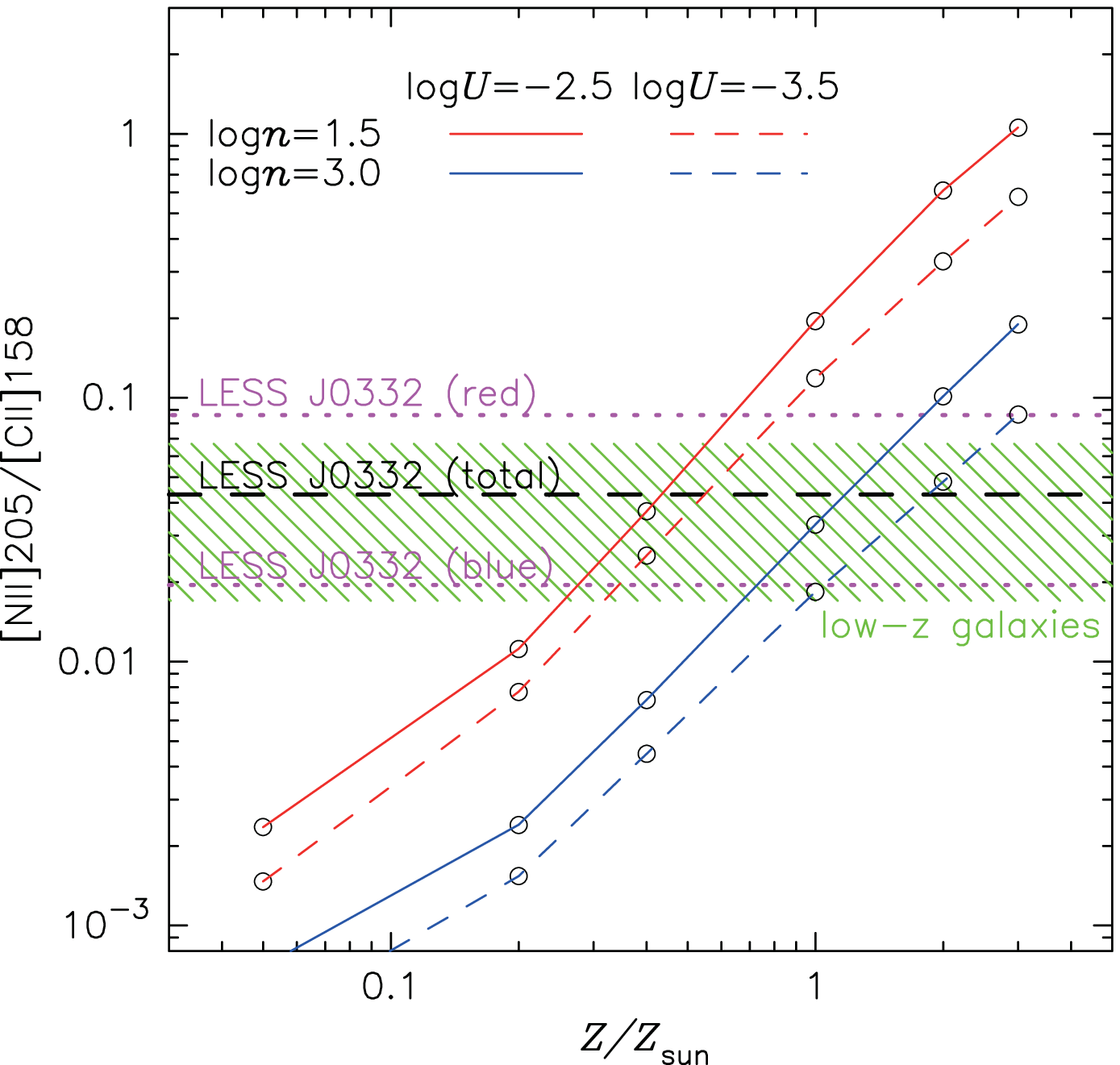}
\caption{
  Observed [N~{\sc ii}]/[C~{\sc ii}] flux ratios compared with model 
  predictions. The green hatched range denotes the observed range for 
  low-$z$ galaxies. The horizontal dashed line shows our ALMA result on 
  LESS J0332, where the emission-line fluxes are measured by 
  integrating the best-fit Gaussian function. Dotted magenta lines denote 
  the flux ratio at the red part (from --350 km s$^{-1}$ to --150 km s$^{-1}$) 
  and blue part (from --550 km s$^{-1}$ to --350 km s$^{-1}$) of the lines.
  The red and blue lines show Cloudy model results as a function of 
  $Z_{\rm gas}$ with $\log n_{\rm HII}$ = 1.5 and 3.0 respectively, while 
  solid and dashed lines denote the models with $\log U_{\rm HII}$ = 
  --2.5 and --3.5, respectively.
}
\end{figure}

To explore the gas metallicity in LESS J0332 more quantitatively, we carried
out model calculations using Cloudy \citep{1998PASP..110..761F} version 
08.00. Since the [C~{\sc ii}] line arises in both H~{\sc ii} 
regions and PDRs, a consistent treatment to connect those two regions is 
required to investigate the [N~{\sc ii}]/[C~{\sc ii}] flux ratio. We assumed a 
pressure-equilibrium gas cloud that is characterized by certain gas densities
and ionization parameters at the illuminated face ($n_{\rm HII}$ and 
$U_{\rm HII}$) for each model run. Here we examine gas clouds with 
$\log n_{\rm HII} =$ 1.5 and 3.0, and $\log U_{\rm HII} =$ --3.5 and --2.5, 
for $Z_{\rm gas}/Z_\odot =$ 0.05 -- 3.0. Note that the gas density in PDRs is 
higher than $\log n_{\rm HII}$ under the pressure-equilibrium assumption. 
The input continua are the starburst99 constant star-formation model 
spectra \citep{1999ApJS..123....3L} with an age of 1 Myr and a stellar 
metallicity equal to the gas metallicity. We did not take AGN effects into 
account in the models since the expected AGN contribution is small (see 
\S3.1). The relative chemical composition of gas clouds is scaled from the 
solar elemental abundance ratios except for nitrogen, which scales as 
$Z_{\rm gas}^2$ at $Z_{\rm gas} > 0.23 Z_\odot$ \citep[see][]
{2002ApJS..142...35K}. Orion-type graphite and silicate grains are 
included. Calculations are stopped at the depth of $A_{\rm V} = 100$ mag 
to cover the whole [C~{\sc ii}]\,158\,$\mu$m emitting regions, following
\citet{2005ApJS..161...65A}. Note that the resulting [N~{\sc ii}]/[C~{\sc ii}] 
flux ratios are not sensitive to the adopted stopping criterion. The calculation 
details are given in \citet{2011A&A...526A.149N}. 

The [N~{\sc ii}]/[C~{\sc ii}] flux ratio predicted by the model runs are 
compared with observations in Fig.~3. The models show that the 
[N~{\sc ii}]/[C~{\sc ii}] flux ratio increases monotonically with $Z_{\rm gas}$. 
The moderate dependences on parameters other than $Z_{\rm gas}$, 
such as $n_{\rm HII}$ and $U_{\rm HII}$, prevent us from determining the 
accurate $Z_{\rm gas}$ value. The observed [N~{\sc ii}]/[C~{\sc ii}] flux 
ratio ($\sim 0.043$) infers $\log (Z_{\rm gas}/Z_\odot) \sim 0.0 \pm 0.3$; 
i.e., consistent to the solar metallicity. Although the constraint on the gas 
metallicity is currently fairly loose, by observing additional 
lines in the future (such as [N~{\sc ii}]\,122\,$\mu$m and [O~{\sc i}]\,145\,$\mu$m) it 
will be possible to constrain the gas properties (density and ionization 
parameter) and therefore determine the gas metallicity much more 
accurately.

However, we note that the velocity profiles of the [N~{\sc ii}]\,205\,$\mu$m 
and [C~{\sc ii}]\,158\,$\mu$m emission are different (Fig.~2). In the redder 
part of the line the [N~{\sc ii}]/[C~{\sc ii}] flux ratio is significantly lower than 
that in the bluer part. If the different [N~{\sc ii}]/[C~{\sc ii}] ratios for different 
velocities are caused by metallicity variations, this implies that the 
system is not chemically homogeneous, but probably resulting from a 
merging systems where gas with different metallicities has not fully mixed 
yet. More specifically, the redder component with a high 
[N~{\sc ii}]/[C~{\sc ii}] flux ratio may be associated with a chemically 
enriched galaxy that is merging with a metal poor galaxy, associated with 
the bluer component characterized by the low [N~{\sc ii}]/[C~{\sc ii}] ratio.
To investigate this scenario more quantitatively, we measured the 
[N~{\sc ii}]/[C~{\sc ii}] flux ratio on the redder part (--350 km s$^{-1}$ 
$< V <$ --150 km s$^{-1}$) and on the bluer part (--550 km s$^{-1}$ 
$< V <$ --350 km s$^{-1}$) of the lines, resulting in a flux ratio of 0.086 and 
0.020, respectively. These values roughly correspond to the highest and 
lowest flux ratios seen in nearby galaxies (Fig.~3), which is indicative of 
solar--supersolar metallicity in the former case and sub-solar metallicity 
in the later case. Note that the two distinct [N~{\sc ii}]\,205\,$\mu$m 
velocity components are seen in another high-$z$ [N~{\sc ii}]\,205\,$\mu$m 
emitter, HLS J091828.6+514223 (Combes et al. 2012), which may be another 
example of a chemically inhomogeneous system at high-$z$.

The low-metallicity component in LESS J0332 was already 
identified in our previous work on this SMG, where we argued (based at 
that time solely on the [C~{\sc ii}]\,158\,$\mu$m, FIR and CO properties)
that the gas metallicity in LESS J0332 is low \citep{2011A&A...530L...8D}. 
However, the most interesting result obtained here is that this system 
does also host a significant metal-rich component (with solar--supersolar 
metallicity), indicating that this system is chemically highly evolved 
already at $z = 4.76$ (the cosmic age of 1.27 Gyr).

This result is consistent with past metallicity studies on high-$z$ 
AGNs. The diagnostic flux ratios of some UV metallic permitted lines of 
type-1 quasars show no redshift evolution of the 
quasar broad-line region up to $z \sim 6$ \citep[e.g.,][]
{2006A&A...447..157N,2009A&A...494L..25J}. The metallicity in narrow-line 
regions of type-2 AGNs (which trace larger spatial scales 
and are therefore more closely related to the host galaxy properties than to the 
broad-line region) also show no 
redshift evolution \citep[e.g.,][]{2006A&A...447..863N} even up to 
$z \sim 5.2$ \citep{2011A&A...532L..10M}. Note that the high-$z$ AGNs 
investigated in those studies are hosted mostly by massive 
galaxies \citep[see, e.g.,][]{2007ApJS..171..353S}, similar to SMGs 
\citep[e.g.,][]{2005ApJ...635..853B}. These results on the 
metallicity of AGNs and LESS J0332 suggest that the chemical 
evolution of massive systems has progressed very rapidly in the early 
epoch, $z>5$, which is qualitatively consistent with the so-called downsizing 
evolution seen in galaxies and AGNs \citep[e.g.,][]{2006MNRAS.366..499D,
2011ApJ...728L..25I}.

\begin{acknowledgements}
This paper makes use of the following cycle 0 ALMA data:
ADS/JAO.ALMA\#2011.0.00268.S (PI: T. Nagao). ALMA is a partnership of 
NINS (Japan), ESO (representing its member states), and NSF (USA), 
together with NRC (Canada) and NSC and ASIAA (Taiwan), in cooperation 
with the Republic of Chile. The Joint ALMA Observatory is operated by NAOJ, 
ESO, and AUI/NRAO.
We thank the ALMA staff for their supports, and Fabian Walter and the
anonymous referee for useful 
comments. Cloudy was developed and
publicly released by G. Ferland and his collaborators. T.N. is financially 
supported by JSPS (grant no. 23654068). B.H. is a JSPS fellow.
\end{acknowledgements}

\bibliographystyle{aa}

\end{document}